\documentclass[a4paper,fleqn,10pt]{article}
\pdfoutput=1
\usepackage{amsmath}
\usepackage{amssymb}
\usepackage{array}
\usepackage{calc}
\usepackage{longtable}
\usepackage{multirow}
\usepackage{pstricks}
\usepackage{graphicx}
\usepackage{xspace}
\usepackage{units}

\numberwithin{equation}{section}
\usepackage{mciteplus}
\usepackage[a4paper,pdfborder={0 0 0}]{hyperref}
\usepackage[format=hang,labelfont=bf,hypcap=true]{caption}
\usepackage{subcaption}
\usepackage{sectsty}
\allsectionsfont{\sffamily}
\subsubsectionfont{\mdseries\itshape\large}
\makeatletter
\DeclareRobustCommand*{\bfseries}{%
  \not@math@alphabet\bfseries\mathbf
  \fontseries\bfdefault\selectfont
  \boldmath
}
\makeatother
\let\spreprint\empty
\newcommand{\preprint}[1]{\def\spreprint{\protect#1}}
\let\sinstitute\empty
\newcommand{\institute}[1]{\def\sinstitute{\protect#1}}
\makeatletter
\renewcommand{\maketitle}{\begingroup
  \null\thispagestyle{empty}%
    \ifx\spreprint\empty
      \vskip 5ex
    \else
      \flushright\large\spreprint\vskip 2ex
    \fi
    \vskip 5ex
    \flushleft
      {\sffamily\bfseries\huge\@title}\vskip 6ex
      \@author\vskip 2ex
      \ifx\sinstitute\empty
      \else
        {\small\sinstitute}
      \fi
    \vskip 5ex
  \endgroup
}
\makeatother
\renewenvironment{abstract}{\begin{center}
  {\large\sffamily\bfseries Abstract: }
  \begin{minipage}[t]{0.75\textwidth}
}{\end{minipage}\end{center}\vskip 10ex}

\numberwithin{equation}{section}
\newcommand{\MCatNLO}{M\protect\scalebox{0.8}{C}@N\protect\scalebox{0.8}{LO}\xspace}

\newcommand{\POWHEG}{P\protect\scalebox{0.8}{OWHEG}\xspace}

\newcommand{\MEPS}{M\scalebox{0.8}{E}P\scalebox{0.8}{S}@LO\xspace}

\newcommand{\MEPSatNLO}{M\scalebox{0.8}{E}P\scalebox{0.8}{S}@NLO\xspace}

\newcommand{\Pythia}{P\protect\scalebox{0.8}{YTHIA}\xspace}

\newcommand{\GoSam}{G\protect\scalebox{0.8}{O}S\protect\scalebox{0.8}{AM}\xspace}
\newcommand{\Golem}{G\protect\scalebox{0.8}{OLEM}95\xspace}
\newcommand{\Samurai}{S\protect\scalebox{0.8}{AMURAI}\xspace}

\newcommand{\Sherpa}{S\protect\scalebox{0.8}{HERPA}\xspace}
\newcommand{\Comix}{C\protect\scalebox{0.8}{OMIX}\xspace}

\newcommand{\Amegic}{A\protect\scalebox{0.8}{MEGIC++}\xspace}

\long\def\symbolfootnote[#1]#2{\begingroup%
\def\thefootnote{\fnsymbol{footnote}}\footnote[#1]{#2}\endgroup}

\newcommand{\rbr}[1]{\left( #1\right)}
\newcommand{\abr}[1]{\langle #1\rangle}

\newcommand{\sbr}[1]{\left[ #1\right]}
\newcommand{\im}{\imath}
\newcommand{\jm}{\jmath}

\newcommand{\done}{{\rm d}}

\newcommand{\mc}[1]{\mathcal{#1}}
\newcommand{\mr}[1]{\mathrm{#1}}

\newcommand{\dst}{\displaystyle}

\newcommand{\bea}{\begin{eqnarray}}
\newcommand{\eea}{\end{eqnarray}}
\newcommand{\bi}{\begin{itemize}}
\newcommand{\ei}{\end{itemize}}

\newcommand{\trule}{\rule[-1.5mm]{0mm}{6mm}}

\hypersetup{
  pdfauthor={Stefan Hoeche, Junwu Huang, Gionata Luisoni,
  Marek Schoenherr, Jan-Christopher Winter},
  pdftitle={Zero and one jet combined NLO analysis
    of the top quark forward-backward asymmetry}
}
\preprint{{\small SLAC-PUB 15553\\MPP-2013-154\\
  IPPP/13/39\\DCPT/13/78\\MCNET/13/07\\LPN13-035\\}}
\author{Stefan H{\"o}che$^1$, Junwu Huang$^1$, 
  Gionata Luisoni$^2$, Marek Sch{\"o}nherr$^3$,
  Jan Winter$^2$}
\title{Zero and one jet combined NLO analysis\\[2mm]
  of the top quark forward--backward asymmetry}
\institute{$^1$ SLAC National Accelerator Laboratory, 
  Menlo Park, CA 94025, USA\\
  $^2$ Max-Planck Institut f{\"u}r Physik,
  F{\"o}hringer Ring 6, 80805 M{\"u}nchen, Germany\\
  $^3$ Institute for Particle Physics Phenomenology,
  Durham University, Durham DH1 3LE, UK\\}
\begin{document}
\maketitle
\begin{abstract}
  We present an analysis of the forward--backward asymmetry in the production
of top quark pairs at the Tevatron collider. We use novel Monte Carlo methods
for merging matrix elements and parton showers to combine NLO QCD predictions
for $t\bar{t}$ and $t\bar{t}$+jet production. Theoretical uncertainties are 
quantified in detail. We find agreement with experimental data on the
transverse momentum dependence of the asymmetry.

\end{abstract}
\section{Introduction}
\label{sec:intro}

The forward--backward asymmetry in the production of top quark pairs
offers great opportunities to study the physics both within and beyond
the Standard Model (SM). At $p\bar{p}$ colliders,
the asymmetry in dependence on the observable $O$ is defined as
\begin{equation}
  A_\mathrm{FB}(O)=\frac{\done\sigma_{t\bar{t}}/\done O|_{\Delta y>0}
    -\done\sigma_{t\bar{t}}/\done O|_{\Delta y<0}}{
    \done\sigma_{t\bar{t}}/\done O|_{\Delta y>0}
    +\done\sigma_{t\bar{t}}/\done O|_{\Delta y<0}}\;,
\end{equation}
where $\Delta y=y_t-y_{\bar t}$ is the rapidity difference 
between the top and the antitop quark. 

Unexpectedly large inclusive and differential 
asymmetries were found in various measurements at 
the Tevatron~\cite{Abazov:2007ab,*Aaltonen:2008hc,
  *Aaltonen:2011kc,*Abazov:2011rq}. By now, both the
CDF and D\O\ collaborations observed values that cannot
be described by predictions based on the Standard
Model~\cite{Kuhn:1998kw,*Antunano:2007da,*Almeida:2008ug,*Kidonakis:2011zn,
  *Ahrens:2011uf,*Kuhn:2011ri,*Alioli:2011as,*Melnikov:2011qx,*Manohar:2012rs,
  *Brodsky:2012ik,Dittmaier:2008uj,Campbell:2012uf}.
The CDF collaboration has reported on forward--backward asymmetries 
at $\sqrt{s} = 1.96$\:TeV using the full Run II data set~\cite{Aaltonen:2012it}.
Their result was compared to theoretical predictions from various Monte Carlo 
event generators and next-to-leading order (NLO) calculations in the
Standard Model.
This analysis confirmed a discrepancy between theory and experiment which was 
observed earlier. It is most significant for those $t\bar{t}$ events
with large invariant mass, $m_{t \bar{t}}\geq450$\:GeV. The
inclusive parton-level asymmetry was measured as $0.164\pm0.047$
considering all pair masses, and $0.295\pm0.067$ 
for $m_{t\bar t}\geq450$\:GeV.
This needs to be compared to theoretical predictions from various 
event generators in the unrestricted and
high $t\bar t$ mass region: $0.067\pm0.020$ and $0.089\pm0.027$ 
from \MCatNLO~\cite{Frixione:2002ik,Frixione:2003ei},
$0.066\pm0.020$ and $0.100\pm0.030$ from
\POWHEG~\cite{Nason:2004rx,*Frixione:2007vw,
  *Alioli:2010xd,*Hoeche:2010pf,Frixione:2007nw},
and $0.073\pm0.022$ and $0.110\pm0.033$ from
MCFM~\cite{Campbell:2012uf,Campbell:2010ff}.
In all cases, electroweak corrections had been applied.
The linear behavior of $A_\mathrm{FB}$ with increasing $\Delta y$
and $m_{t\bar t}$ persists, but the prediction for the slope
is reduced by $\sim 2\,\sigma$ compared to earlier measurements 
for both the $m_{t\bar t}$ and the $\Delta y$ dependence.

The observation of a large asymmetry has triggered substantial theoretical 
investigation. Various new physics models have been proposed to explain 
the discrepancy seen within the SM, such as $t\bar t$ production via a heavy 
axial color octet or a flavor changing $Z'$ boson~\cite{Frampton:1987dn,
  *Agashe:2006hk,*Barger:2006hm,*Frederix:2007gi,*Djouadi:2009nb,
  *Bauer:2010iq,*Choudhury:2010cd,*Cao:2010nw,
  *Barcelo:2011fw,*Barcelo:2011vk,*Kamenik:2011wt,*AguilarSaavedra:2011ci,
  *Tavares:2011zg,*Berger:2011hn,*Barger:2011ya,*Bai:2011ed,
  *AguilarSaavedra:2012va,*Alvarez:2012ca,*Drobnak:2012cz,*Drobnak:2012rb,
  *DaRold:2012sz,*Han:2012dd,*Palle:2012kma,*Huang:2012rh,
  *Baumgart:2013yra,*Li:2013dia,*Chivukula:2013kw,*Arbuzov:2013fa,*Guo:2013dc}.
However, in order to ensure that the asymmetry is indeed a first hint of
new physics beyond the Standard Model, a systematic study of
QCD and Electroweak (EW) corrections at NLO and beyond must be performed 
to reduce theoretical uncertainties as much as possible. 
It was pointed out in~\cite{Skands:2012mm} that color flows from
incoming quarks to the top quark and from antiquarks to the antitop
quark lead to more radiation
when the top quark goes backward. This generates a positive asymmetry
already at the level of parton showers that include color coherence effects.
NLO QCD predictions for $t\bar{t}$ and $t\bar{t}+$jet~\cite{
  Dittmaier:2009jq,Melnikov:2010iu} exhibit a non-constant $K$-factor,
such that additional effects are expected.
Much attention has also been paid to the calculation of the EW 
contributions to the asymmetry from pure EW interactions 
and the interplay between EW and QCD
processes~\cite{Hollik:2011ps,*Kuhn:2013zoa}.
A combined correction of $26\%$ on top of the QCD prediction was determined 
at $O(\alpha_{s}^2\alpha)$ and $O(\alpha^2)$. Tremendous efforts have
been recently made to complete the full NNLO QCD cross section
calculation~\cite{Baernreuther:2012ws,*Czakon:2012zr,
  *Czakon:2012pz,*Czakon:2013goa}.
Soft-gluon resummation was also performed in this context~\cite{Beneke:2009rj,
  *Czakon:2009zw,*Ahrens:2011px,*Kidonakis:2011ca,*Beneke:2011mq,
  *Cacciari:2011hy,*Moch:2012mk}.

It is important to note, that all general-purpose Monte Carlo 
event generators which are currently being used by experiments provide 
at most the inclusive prediction for $t\bar{t}$ production at the NLO 
matched to a parton shower~\cite{Frixione:2003ei,Frixione:2007nw}.
While calculations of $t\bar{t}$+jet production at NLO have been matched
to parton showers independently~\cite{Kardos:2011qa}, they have not yet 
been combined with the inclusive simulation of $t\bar{t}$ production in 
a manner that allows for improved predictions of $A_{\rm FB}$.
We remedy this situation in the present publication, providing
a merged simulation of $t\bar{t}$ and $t\bar{t}+$jet production
at hadron colliders, which preserves both the NLO accuracy of the
fixed-order prediction and the logarithmic accuracy of the parton shower.
We are thus able to make predictions for both, the transverse momentum 
dependent asymmetry above a certain threshold and the inclusive asymmetries, 
which depend strongly on both, real and virtual higher-order QCD corrections. 
We do not include electroweak corrections in this publication, 
these can be inferred from~\cite{Hollik:2011ps}.

We employ the \MEPSatNLO technique for combining multiple NLO
parton-level calculations with parton showers. The method was
introduced in~\cite{Gehrmann:2012yg,*Hoeche:2012yf} and is implemented
in the general-purpose event generator 
\Sherpa~\cite{Gleisberg:2003xi,*Gleisberg:2008ta}. Virtual corrections
are computed using the \GoSam~\cite{Cullen:2011ac,*Cullen:2011xs} package,
which makes use of the program \Samurai~\cite{Mastrolia:2010nb} based
on integrand reduction techniques~\cite{Ossola:2006us,*Ellis:2007br},
and the tensor integral library \Golem~\cite{Binoth:2008uq,*Cullen:2011kv}.
The interface between \Sherpa and \GoSam~\cite{LSTW} uses the
Binoth--Les--Houches accord (BLHA)~\cite{Binoth:2010xt}.

A fair amount of uncertainty is involved in parton-shower simulations 
of $A_\mathrm{FB}$, both inclusive and differential
$A_\mathrm{FB}$~\cite{Skands:2012mm}.
Some of these uncertainties will be eliminated by a combination of
higher-multiplicity NLO calculations with the inclusive result.
Some of them remain, such as the uncertainty related to the
choice of exponent in the Sudakov factor of the parton shower. 
This has been discussed extensively in~\cite{Hoeche:2011fd}.
We do not attempt to systematically improve the parton shower here.
Therefore, our ability to describe the inclusive asymmetry is 
still somewhat limited. However, we can quantify the possible
impact of a systematic improvement at higher parton multiplicity
by judging the impact of matrix-element plus parton-shower merging 
at the NLO. Moreover, we readily provide an NLO-accurate 
prediction for the transverse momentum dependent asymmetry 
for all but the first bin in $p_{T,t\bar{t}}$.~\footnote{
  Note that this is a major difference between our results
  and the predictions from~\cite{Frederix:2012ps}.
  Other differences include the treatment of color
  (cf.\ Sec.~\ref{sec:mcatnlo}) and truncated shower
  effects (cf.\ Sec.~\ref{sec:mepsatnlo}).}

The paper is organized as follows: Section~\ref{sec:mcatnlo}
introduces the \MCatNLO method, as implemented in \Sherpa,
and discusses color-coherence effects on $A_{\rm FB}$. 
Section~\ref{sec:mepsatnlo} briefly 
reviews the \MEPSatNLO technique and discusses related uncertainties.
Section~\ref{sec:results} presents our final predictions, and
Sec.~\ref{sec:conclusions} contains some concluding remarks.

\section{\texorpdfstring{\protect\MCatNLO}{MC@NLO} for massive particles}
\label{sec:mcatnlo}

The \MCatNLO method is a modified subtraction scheme, which relies on 
the unitarity condition of the parton shower. Virtual corrections are
approximated by the parton shower as the counterpart of real-emission
corrections, integrated over the phase space of the emission.
This implies that parton showers do not change the weight of a Monte
Carlo event.
They simply move events from the $n$-parton phase space to the 
$(n+1)$-parton phase space by means of branching processes.

Parton branching as implemented in \MCatNLO can be described by the 
following equation, which determines the expectation value of an arbitrary, 
infrared-safe observable, denoted by $O$,
\begin{equation}\label{eq:mcatnlo}
  \abr{O}\,=\;\int\done\Phi_B\,\bar{\mr{B}}^{\rm(A)}(\Phi_B)\,
    \mc{F}^{\rm(A)}(\mu_Q^2,O)
  +\int\done\Phi_R\,\mr{H}^{\rm(A)}(\Phi_R)\,\mc{F}_1(t,O)\;.
\end{equation}
In this context,
\begin{equation}\label{eq:gen_mcatnlo}
  \mc{F}^{\rm(A)}(\mu_Q^2,O)\,=\;
    \Delta^{\rm(A)}(t_c,\mu_Q^2)\,O(\Phi_B)
    +\int_{t_c}^{\mu_Q^2}\!\done\Phi_1\,
      \frac{\mr{D}^{\rm(A)}(\Phi_B,\Phi_1)}{\mr{B}(\Phi_B)}\,
      \Delta^{\rm(A)}(t,\mu_Q^2)\,\mc{F}_1(t,O)
\end{equation}
is the generating functional of the \MCatNLO, while 
$\mc{F}_n(t,O)$ denotes the generating functional 
of the parton shower. $\Phi_B$ and $\Phi_R$ denote the 
Born- and real-emission phase space, and $\Phi_1$ is the
phase space associated with the emission of an additional parton,
i.e.\ $\done\Phi_R=\done\Phi_B\cdot\done\Phi_1$. 
It is parametrized in the standard manner as 
$\done\Phi_1=\done t\,\done z\,\done\phi\,J(t,z)$, where $t$ 
is called the evolution variable, $z$ is called the splitting variable, 
$\phi$ is an azimuthal splitting angle and $J(t,z)$ is a Jacobian factor.
Thus, where appropriate, $t\equiv t(\Phi_1)$ is understood.
The functions $\bar{\mr{B}}^{\rm(A)}$ and $\mr{H}^{\rm(A)}$ are called
the NLO-weighted Born differential cross section and the hard remainder
function, defined as
\begin{equation}\label{eq:mcatnlo_bbar_h}
  \begin{split}
  \bar{\mr{B}}^{\rm(A)}(\Phi_B)\,=&\;\mr{B}(\Phi_B)+\tilde{\mr{V}}(\Phi_B)
    +\mr{I}^{\rm(S)}(\Phi_B)+\int\done\Phi_1\,\Big[\,
      \mr{D}^{\rm(A)}(\Phi_B,\Phi_1)\,\Theta\left(\mu_Q^2-t\right)
      -\mr{D}^{\rm(S)}(\Phi_B,\Phi_1)\,\Big]\;,\\
  \mr{H}^{\rm(A)}(\Phi_R)\,=&\;\mr{R}(\Phi_R)
  -\mr{D}^{\rm(A)}(\Phi_R)\,\Theta\left(\mu_Q^2-t\right)\;.
  \end{split}
\end{equation}
The terms $\mr{B}$ and $\mr{R}$ represent Born- and real-emission 
matrix elements, including flux and parton luminosity factors;
$\mr{D}^{\rm(S)}$ and $\mr{I}^{\rm(S)}$ are the subtraction and
integrated subtraction terms, respectively.
$\tilde{\mr{V}}$ represents the virtual corrections, including collinear
mass-factorization counterterms. $\mr{D}^{\rm(A)}$ is the resummed part 
of the real-emission correction, which must approach $\mr{R}$ in both 
the collinear and the soft limit.

Within the event generator \Sherpa, 
$\mr{D}^{\rm(A)}$ is defined by the dipole subtraction terms employed
in the method of Catani and Seymour (CS)~\cite{Catani:1996vz}. Their
phase space is restricted by the resummation scale 
$\mu_Q^2$~\cite{Hoeche:2011fd}. The corresponding dipole insertion 
operators are modified such that their helicity summed splitting 
operator is positive definite, while negative values induced through 
spin dependence and color insertion operators are kept. 
Thus, spurious negative terms arising from arbitrary finite corrections 
are not resummed through $\mr{D}^{\rm(A)}$.

So far, the \MCatNLO method had been implemented only for massless partons
in \Sherpa. In the context of this work, we extended the implementation
to massive partons, using kinematics and phase-space factorization in the 
method of Catani, Seymour, Dittmaier and Trocsanyi (CDST)~\cite{Catani:2002hc}. 
The evolution variable is chosen to be a Lorentz invariant transverse momentum.
Using the definitions of \cite{Catani:1996vz}, for final-state branchings 
$\{\widetilde{\im\jm},\tilde{k}\}\to\{i,j,k\}$ we have
(denoting parton masses by $m$)
\begin{equation}\label{eq:def_kt_fss}
  t^{\rm (FS)}\,=\;2\,p_ip_j\,\tilde{z}_{i,jk}(1-\tilde{z}_{i,jk})
    -(1-\tilde{z}_{i,jk})^2\,m_i^2-\tilde{z}_{i,jk}^2\,m_j^2\;,
\end{equation}
while for initial-state branchings
$\{\widetilde{a\jm},\tilde{k}\}\to\{a,j,k\}$ we use
\begin{equation}\label{eq:def_kt_iss}
  t^{\rm (IS)}\,=\;2\,p_ap_j\,(1-x_{aj,k})\;.
\end{equation}
Note that these definitions are independent of the type of the
spectator parton, and they are also used in the parton shower.

\begin{figure}[t!]
  \centering
  \includegraphics[width=0.495\textwidth]{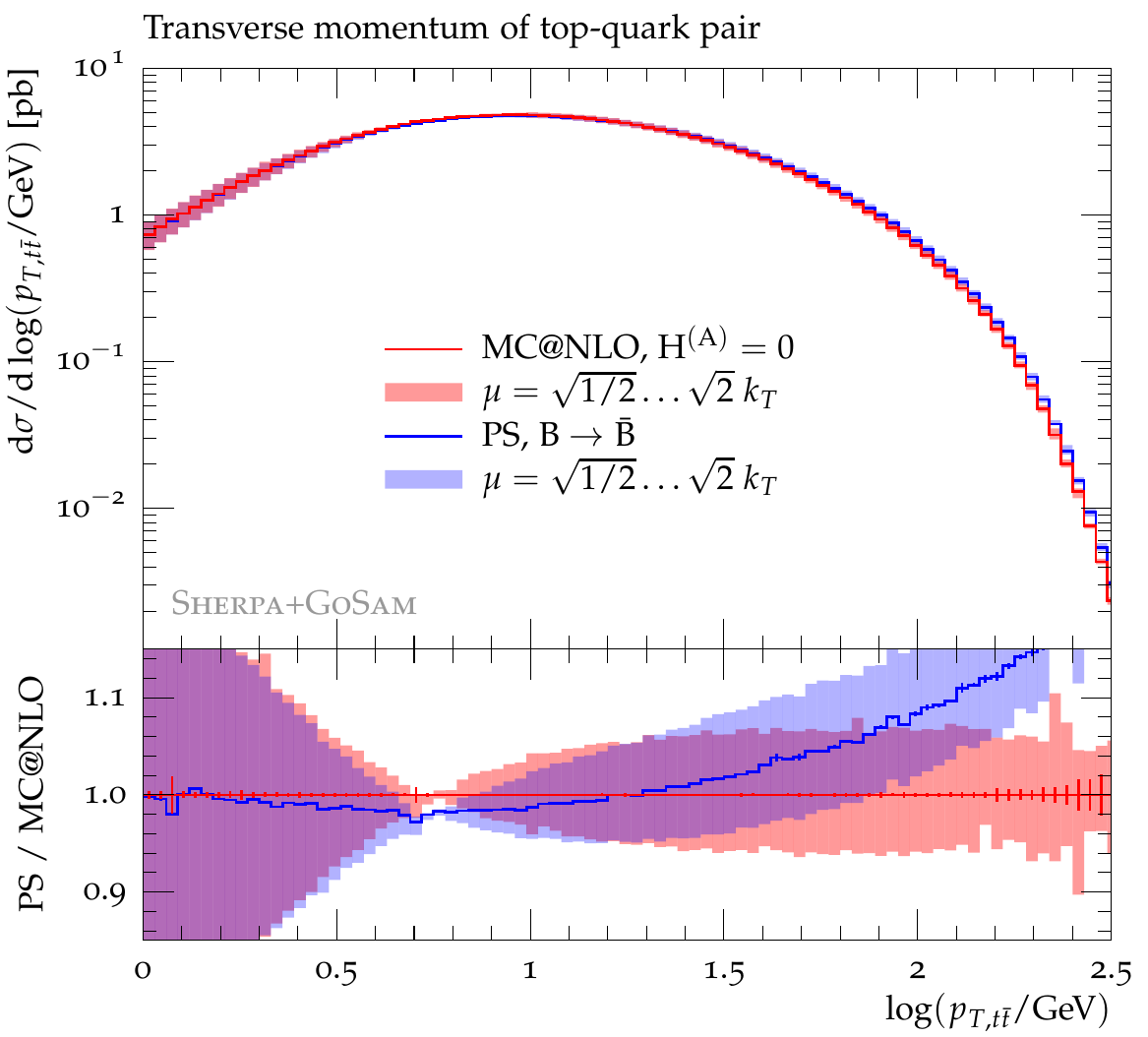}\hfill
  \includegraphics[width=0.495\textwidth]{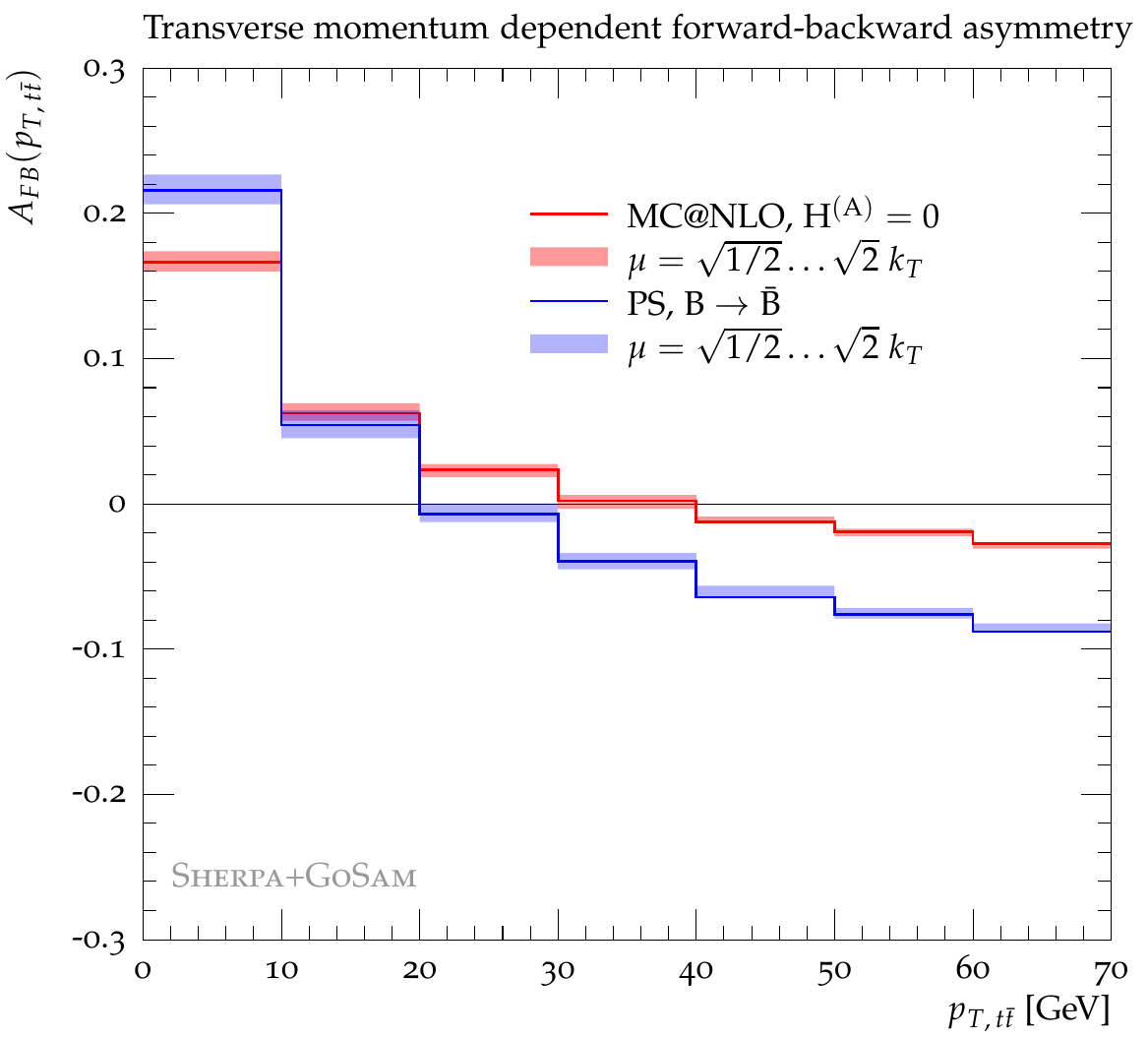}
  \caption{\label{fig:ptt_coherence}%
    Transverse momentum spectrum of the $t\bar{t}$ pair (left)
    and $p_T$-dependent forward--backward asymmetry (right).
    We compare the \MCatNLO prediction (red) and the parton-shower
    result (blue).
    Hard remainder terms have been set to zero in the \MCatNLO simulation,
    while the parton shower has been reweighted with the local $K$-factor 
    $\bar{\mr{B}}^{\rm(A)}/\mr{B}$ in order to make the two results comparable.
    Uncertainty bands stem from varying the scale of strong couplings in the
    resummation.}
\end{figure}

The generating functional of the corresponding parton shower, which was
described in~\cite{Schumann:2007mg} is given by
\begin{equation}\label{eq:gen_ps}
  \mc{F}_n(t,O)\,=\;\sum_{l=n}^\infty
    \prod_{i=n+1}^l\Bigg[\,\int_{t_c}^{t_{i-1}}\done\Phi_{1,i}\,
      \mr{K}_i(\Phi_{1,i})\,\Delta_{i-1}(t_i,t_{i-1})\Bigg]\,
      \Delta_{l}(t_c,t_l)\,O(\Phi_l)\,\Bigg|_{\,t_n=\,t}
\end{equation}
where $\mr{K}_l$ is the sum of evolution kernels for an $l$-parton state and
$\Delta_n$ is the respective Sudakov factor. We define the parton-shower
cutoff as $t_c$. This parton shower is based on the leading color approximation.
It describes the QCD evolution of processes with only a single color
configuration at leading order particularly well. Such reactions include
the production of jets at $e^+e^-$-colliders or the production of
Drell--Yan lepton pairs at hadron colliders.
Processes with a more complicated color structure at the leading order 
typically pose a problem for any type of parton shower, as the coherent 
emission of soft gluons can only be approximated by angular 
ordering~\cite{Marchesini:1983bm,Marchesini:1987cf}.

This problem is, to some extent, remedied by the implementation of the 
full-color \MCatNLO technique as proposed in~\cite{Hoeche:2011fd}.
We exemplify the corresponding effects on physical observables in
Fig.~\ref{fig:ptt_coherence}. The left panel shows that including 
full color coherence in the first emission has no substantial impact 
on observables like the transverse momentum of the $t\bar{t}$ pair. 
However, the right panel shows that it strongly affects the prediction 
for the $p_T$-dependent forward--backward asymmetry. 

This effect is very different from a typical parton shower uncertainty.
To exemplify this, we also show the effect of changing the scale at which 
the strong coupling is evaluated in the parton shower. Such a variation 
easily generates different transverse momentum spectra, but it does not 
affect the asymmetry, as can be seen by comparing the size of the
red and blue bands in the left and right panels of Fig.~\ref{fig:ptt_coherence}. 
Both bands were generated by varying the
scale in the range $\sqrt{\nicefrac{1}{2}}\,k_T\ldots\sqrt{2}\,k_T$.

Similar statements hold for the choice of the momentum mapping, 
although they apply only within reason. Appendix~\ref{sec:dipole} 
explains how the asymmetry is generated in a parton shower based on 
Catani-Seymour dipole factorization. If the assumption is relaxed 
that the recoil partner of the splitting parton is the color partner 
in the large-$N_c$ limit, then the prediction for the asymmetry will change. 
This has already been demonstrated in~\cite{Skands:2012mm} using 
the \Pythia parton shower.

It should be stressed again that we only include the correct color 
insertion operators for the first emission, all subsequent branchings 
are generated in the standard shower approximation. Nevertheless, 
it is not unreasonable to assume that the overall picture remains 
for a complete full color evolution, as $p_{T,t\bar{t}}$ is largely
generated by the first emission.

\section{Combination of the zero and one jet process}
\label{sec:mepsatnlo}

Most practically implemented methods for combining matrix elements and parton
showers are based on phase space slicing, with the soft part of the phase space 
populated by the parton shower, and the hard part populated by matrix elements, 
either at leading or at next-to-leading order. The slicing parameter is called 
the merging cut, $Q_{\rm cut}$. It is given in a variable referred to as the 
jet criterion, $Q$.

The first working technique to achieve a combination of multiple
NLO calculations for reactions at a hadron collider was introduced 
in~\cite{Gehrmann:2012yg,*Hoeche:2012yf}. It is based on the ideas 
of the CKKW~\cite{Catani:2001cc,*Krauss:2002up,Hoeche:2009rj}
and CKKW-L algorithms~\cite{Lonnblad:2001iq,Lonnblad:2012ix}.
We briefly review this method here to set the stage
for a discussion of its uncertainties.

\subsection{The \texorpdfstring{\protect\MEPSatNLO}{MEPS@NLO} method}

In order to turn the inclusive parton-level calculations
into exclusive $n$-jet predictions, which are to be combined,
one needs to multiply them with no-emission probabilities, 
accounting for the fact that the inclusive cross section 
must be preserved at NLO.

The exclusive contribution to $O$ from a parton-level calculation 
yielding $n$ additional jets compared to the lowest multiplicity process 
reads~\cite{Gehrmann:2012yg,*Hoeche:2012yf}
\begin{equation}\label{eq:nlo_term}
  \begin{split}
    \abr{O}_{n}^{\rm excl}
    \,=&\;\int\done\Phi_{n}\,\Theta(Q(\Phi_{n})-Q_{\rm cut})\;
          \tilde{\mr{B}}_{n}^\text{(A)}(\Phi_n)\,
          \tilde{\mc{F}}^{\rm(A)}_{n}(\mu_Q^2,O\,;<\!Q_{\rm cut})\\
    &+\int\done\Phi_{n+1}\,\Theta(Q(\Phi_{n})-Q_{\rm cut})\,
         \Theta(Q_{\rm cut}-Q(\Phi_{n+1}))\,
         \tilde{\mr{H}}_{n}^\text{(A)}(\Phi_{n+1})\,
         \tilde{\mc{F}}_{n+1}(\mu_Q^2,O\,;<\!Q_{\rm cut})\;,
  \end{split}
\end{equation}
where we have defined the generating functional of a truncated vetoed 
parton shower, $\tilde{\mc{F}}(<\!Q_{\rm cut})$. This parton shower
may generate emissions at each point in a parton shower 
history which corresponds to the matrix-element configuration 
at $\Phi_n$.\footnote{A detailed algorithm for identifying 
  these parton shower histories is discussed 
  in~\cite{Lonnblad:2001iq,Hoeche:2009rj}.}
We describe the respective evolution kernel by summing over all possible 
kernels for the intermediate steps:
\begin{equation}\label{eq:compound_ps_kernel}
   \tilde{\mr{K}}_{n}(\Phi_{1,n+1})\,=\;
   \mr{K}_n(\Phi_{1,n+1})\,\Theta(t_n-t_{n+1})+
   \sum_{i=0}^{n-1}\mr{K}_{i}(\Phi_{1,n+1})\,
      \Theta(t_i-t_{n+1})\,\Theta(t_{n+1}-t_{i+1})\,\Big|_{\,t_0=\mu_Q^2}\;.
\end{equation}
One can now restrict emissions to the appropriate region 
of phase space by replacing $\mr{K}_i(\Phi_{1,n+1})$ $\to$
$\mr{K}_i(\Phi_{1,n+1})$ $\Theta(Q_{\rm cut}-Q(\Phi_i,\Phi_{1,n+1}))$.
This implements the veto procedure. The corresponding generating functional
of the truncated and vetoed parton shower is determined by substituting
$\mr{K}$ with $\tilde{\mr{K}}$ in Eq.~\eqref{eq:gen_ps}. 

We have also defined modified NLO-weighted Born cross sections and
hard remainder functions
\begin{equation}\label{eq:mcatnlon_bbar_h}
  \begin{split}
    \tilde{\rm B}_{n}^{\rm (A)}(\Phi_{n})\,=&\;
      {\rm B}_{n}(\Phi_{n})+\tilde{\mr{V}}_{n}(\Phi_{n})
        +{\rm I}_{n}^{\rm (S)}(\Phi_{n})
      +\int\done\Phi_1\,\sbr{\tilde{\rm D}_{n}^{\rm (A)}(\Phi_n,\Phi_1)
        -{\rm D}_{n}^{\rm (S)}(\Phi_n,\Phi_1)}\\
    \tilde{\mr{H}}_{n}^{\rm(A)}(\Phi_{n+1})\;=&\;
    \mr{R}_{n}(\Phi_{n+1})-
      \tilde{\mr{D}}_{n}^{\rm(A)}(\Phi_{n+1})\\
  \end{split}
\end{equation}
which are given in terms of the compound evolution kernel 
$\tilde{\rm{D}}^{\rm(A)}_n$,
\begin{equation}\label{eq:compound_kernel}
   \tilde{\mr{D}}_{n}^{\rm(A)}(\Phi_{n+1})\,=\;
   \mr{D}_{n}^{\rm(A)}(\Phi_{n+1})\,\Theta(t_{n}-t_{n+1})\;
   +\sum_{i=0}^{n-1}\mr{B}_{n}(\Phi_n)\,\mr{K}_{i}(\Phi_{1,n+1})\,
      \Theta(t_i-t_{n+1})\,\Theta(t_{n+1}-t_{i+1})\,\Big|_{\,t_0=\mu_Q^2}\;.
\end{equation}
The first term on the right-hand side of Eq.~\eqref{eq:compound_kernel} 
resembles the resummed part of the real-emission correction in the original 
\MCatNLO (cf.~Eq.~\eqref{eq:mcatnlo}), while the second term is identical to
the one in Eq.~\eqref{eq:compound_ps_kernel}. It is mandatory to implement 
color coherence in the first term, while it is optional in the second term, 
since the evolution variable is bounded from below by $t_n$.
One finally obtains the generating functional for the combined truncated vetoed 
parton shower plus \MCatNLO, $\tilde{\mc{F}}^{\rm(A)}(<\!Q_{\rm cut})$, by replacing 
the evolution kernel for the first step in $\tilde{\mc{F}}(<\!Q_{\rm cut})$ 
with the compound kernel in Eq.~\eqref{eq:compound_kernel}.

While the structure of Eqs.~\eqref{eq:nlo_term}-\eqref{eq:compound_kernel}
seems quite involved, their interpretation is rather simple:
\MCatNLO itself is a modified subtraction method, which allows to correct
for the mismatch between the parton shower approximation and the full NLO
calculation in the first emission step. As we encounter processes where 
truncated parton showers can generate emissions, we have to take these
emissions into account in the modified subtraction. This leads to the 
definition of Eq.~\eqref{eq:mcatnlon_bbar_h} and the compound evolution
kernel, Eq.~\eqref{eq:compound_kernel}.

\subsection{Theoretical uncertainties}

The uncertainties associated with the above defined \MEPSatNLO method fall 
into three categories: The first are uncertainties related to the choice 
of renormalization and factorization scale. They occur in every NLO
calculation. The second are uncertainties due to the choice of parton
shower parameters, which occur in every parton shower simulation. A typical
example is the choice of resummation scale, $\mu_Q^2$. The last, and final 
uncertainty is related to the choice of the merging cut, $Q_{\rm cut}$, and
the choice of the functional form of jet criterion.\footnote{%
  Strictly speaking this is not an uncertainty, as one would attempt 
  to choose the parameters such that the phase-space region of interest 
  for experimental analyses is always fully covered by respective NLO 
  parton-level calculations~\cite{Catani:2001cc,*Krauss:2002up}.}
We find that the variation associated with the choice of resummation scale
is in most cases smaller than the statistical uncertainty in our simulation.
The other two types of uncertainties are discussed in the following.

\subsubsection*{Merging uncertainties}

\begin{figure}[t!]
  \centering
  \includegraphics[width=0.48\textwidth]{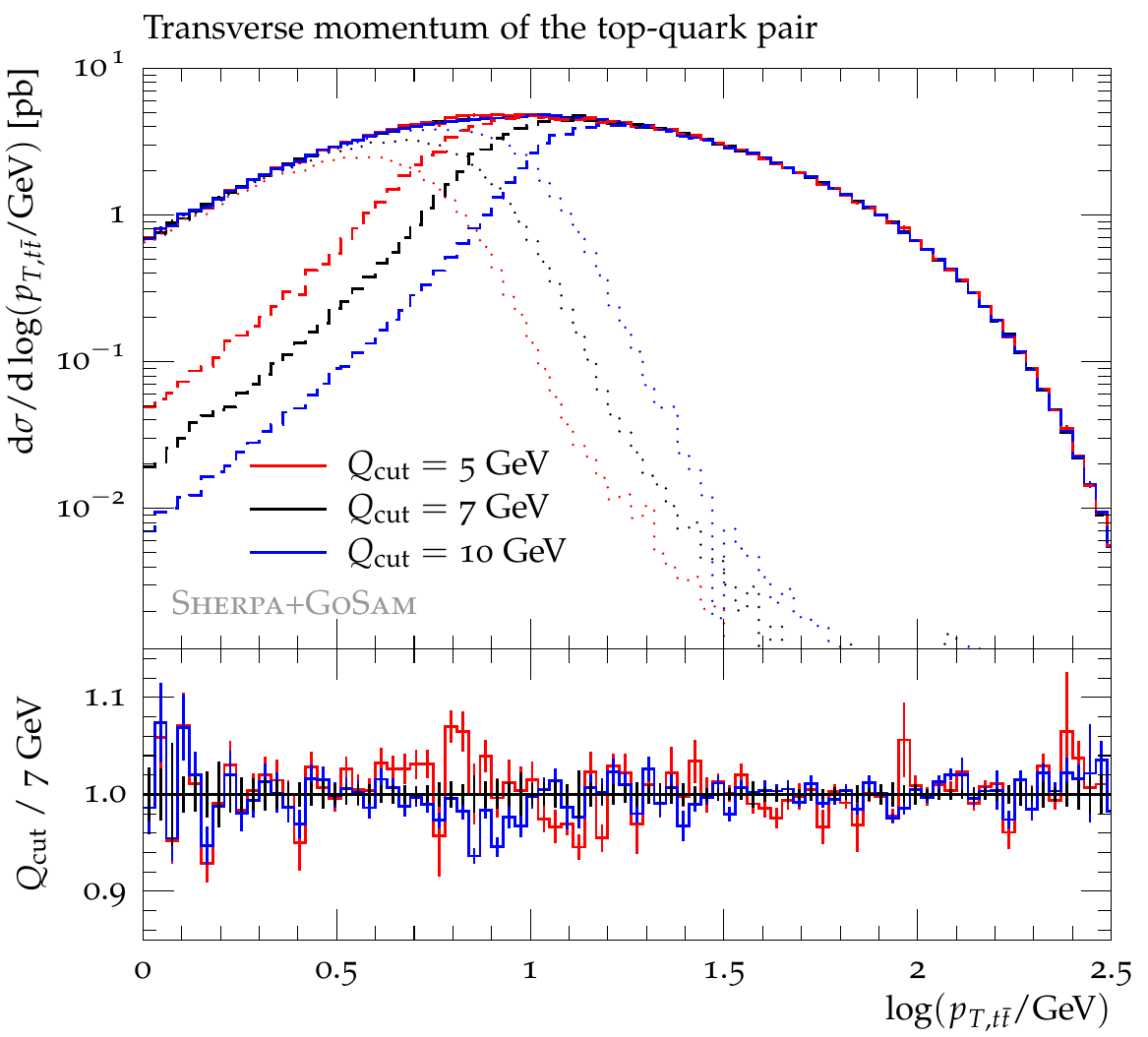}\hfill
  \includegraphics[width=0.48\textwidth]{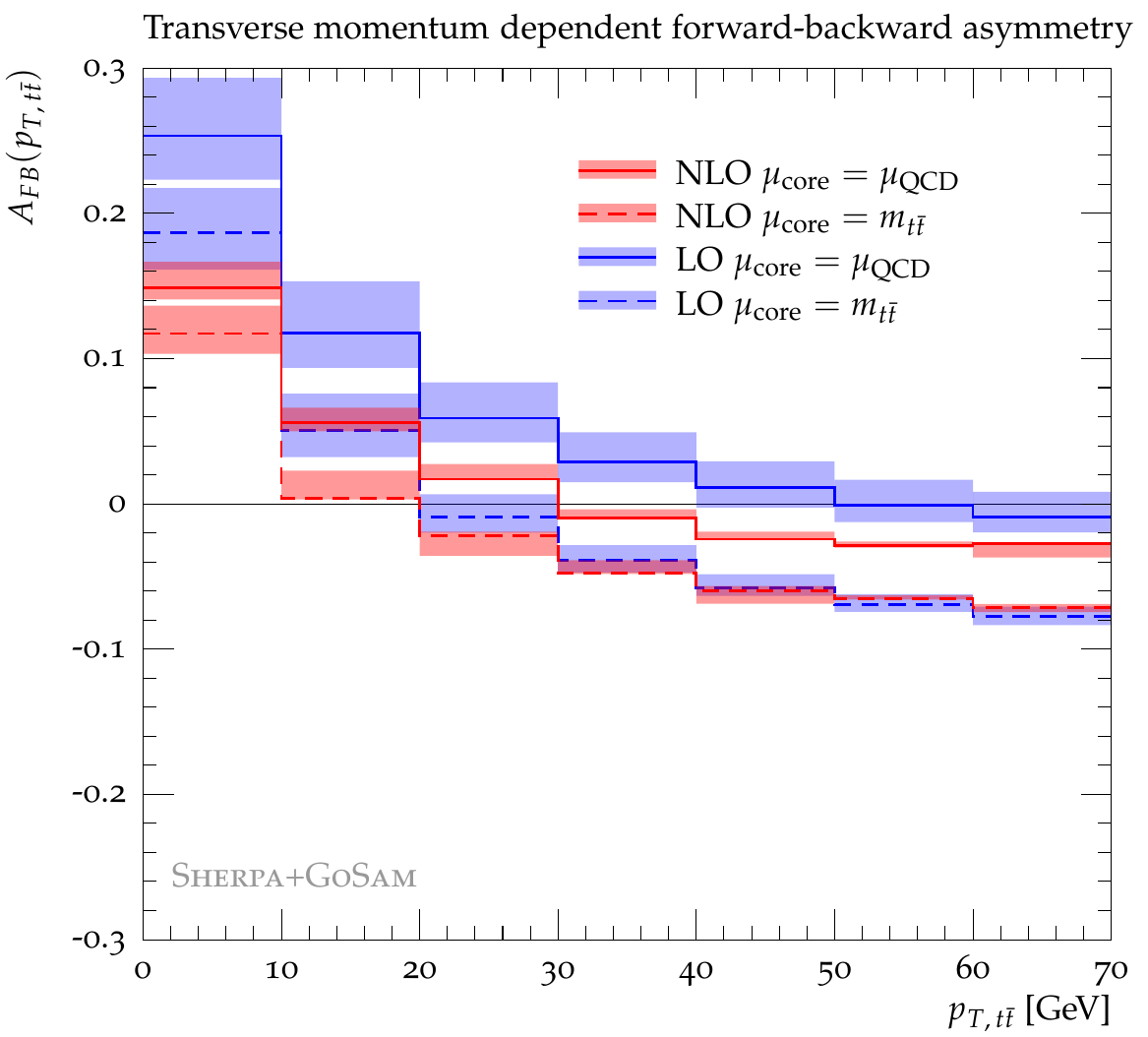}
  \caption{\label{fig:qcut_scale}%
    Systematic uncertainty due to variation of the merging cut (left)
    and due to the scale choice (right). The dotted (dashed) lines 
    in the left panel correspond to contributions from the zero (one)
    jet \MCatNLO. The two bands in the right panel depict results
    from different choices of the functional form of the
    core scale, for more details see the main text. Each band has
    been obtained by varying the respective default scale by factors
    of two.}
\end{figure}

The left panel of Fig.~\ref{fig:qcut_scale} displays the effect of varying
$Q_{\rm cut}$ in the range from $5$ to $10$\:GeV. Effects on the $\log
p_{T,t\bar t}$
spectrum are below 10\%. Potential discontinuities in the transition from the 
zero to the one jet domain are generated by differences between the $t\bar{t}$ 
\MCatNLO at finite transverse momentum and the $t\bar{t}$+jet \MCatNLO.
Out of the two predictions, the $t\bar{t}$ \MCatNLO is less accurate. 
Small discontinuities therefore indicate that it still provides a good estimate
of the $t\bar{t}$+jet production rate at NLO. This means that we can reliably compute
the transverse momentum dependent asymmetry, except for the first bin, where the
prediction is formally still only LO accurate due to the large contribution
from the $t\bar{t}$ \MCatNLO.

Note that we observe unitarity violations in our merging approach.
A comprehensive analysis of the unitarity constraint on the parton shower
in the context of merging algorithms was presented recently~\cite{Lonnblad:2012ng,*Platzer:2012bs}, 
and a new method has been proposed to restore the overall normalization of the
inclusive event sample exactly~\cite{Lonnblad:2012ix}. 
Here, we follow a simpler approach, where unitarity violations may occur,
but their impact on the total cross section is beyond the order at which 
we claim our calculation to be exact~\cite{Gehrmann:2012yg,*Hoeche:2012yf}.

\subsubsection*{Scale uncertainties}
\label{sec:scale_uncertainties}

The right panel of Fig.~\ref{fig:qcut_scale} displays the uncertainty 
arising from a variation of renormalization and factorization scales 
in the range $\nicefrac{1}{2}\,\mu_{R/F}\ldots 2\,\mu_{R/F}$. The two different bands
were generated by choosing the central scale for the $p\bar{p}\to t\bar{t}$
``core'' process in the simulation as either the invariant mass of the
$t\bar{t}$-system, or as twice the product of four-momenta of the 
color-connected partons in the large-$N_c$ limit of the ``core'' process. 
We will refer to the latter scale choice as the ``QCD'' scale.
It is described in detail in Appendix~\ref{sec:qcd_scale}.

Fig.~\ref{fig:qcut_scale} shows that both, the variation of the scale prefactor 
(leading to the uncertainty bands) and the variation of the functional form 
of the core scale (leading to the solid/dashed central histograms) have less 
impact on the predictions from the \MEPSatNLO method than they have on 
predictions from leading--order merging (\MEPS). 
It is interesting to find this effect in an observable like the 
forward--backward asymmetry, where a large fraction of the QCD uncertainties 
are canceled due to taking the ratio between two predictions.

\section{\texorpdfstring{$A_\mathrm{FB}$}{AFB} results}
\label{sec:results}

We now present our $A_\mathrm{FB}$ results generated with the previously 
described techniques. We employ the leading-order matrix element 
generators \Amegic \cite{Krauss:2001iv} and \Comix \cite{Gleisberg:2008fv} 
in conjunction with the automated dipole subtraction provided in \Sherpa 
\cite{Gleisberg:2007md} and the implementation of the Binoth--\-Les Houches 
interface \cite{Binoth:2010xt} to obtain parton-level events at next-to-leading 
order. Virtual matrix elements for $t\bar{t}$ and $t\bar{t}$+jet are provided 
by \GoSam~\cite{Cullen:2011ac,*Cullen:2011xs}. We use a parton shower based 
on Catani--Seymour dipole factorization~\cite{Schumann:2007mg,Hoeche:2009xc} 
and the related \MCatNLO generator~\cite{Hoeche:2011fd,Hoeche:2012fm} 
to generate events at the parton shower level. 

We use \Sherpa version 2.0.0, which includes the modifications described in
Sec.~\ref{sec:mcatnlo}. Parameters are set to their default values, except
for the choice of PDF. We use the MSTW2008 NLO PDF set for \MEPSatNLO and
the MSTW2008 LO set for \MEPS, both with their corresponding 
parametrization of the strong coupling~\cite{Martin:2009iq}. Top quark
decays, multiple interactions and hadronization are not simulated,
since we compare our results to data from the CDF collaboration which
have been corrected to the parton level.

We have validated our parton-level calculations by checking inclusive
cross sections for $t\bar t$+jets against values in the
literature~\cite{Dittmaier:2008uj}. Tests for individual phase-space
points are reported in Appendix~\ref{sec:gosam}. We also verified the
consistency of our results for a number of differential distributions
in inclusive $t\bar t$ production.

\subsection{Inclusive asymmetries}

\begin{table}[t!]
\centering
\begin{tabular}{@{}lccccc@{}}\hline\hline\\[-1mm]
Source & $A_{\rm FB}$ [\%]
& \multicolumn{2}{c}{$A_{\rm FB}(m_{t\bar t})$ [\%]} 
& \multicolumn{2}{c}{$A_{\rm FB}(p_{T,t\bar t})$ [\%]}\\[7pt]
& inclusive
& $m<450$\:GeV & $m>450$\:GeV 
& $p_T<50$\:GeV  & $p_T>50$\:GeV\\[5pt]\hline
\rule[0mm]{0mm}{6mm}%
PRD\textbf{87} (2013) 092002
& $16.4\pm 4.7$
& $8.4\pm 5.5$
& $29.5\pm 6.7$
& $-$
& $-$\\[3mm]
\MEPSatNLO,\ \ {\small$\mu=\mu_\mathrm{QCD}$}
& $8.5$  $^{+0.5}_{-0.5}$ 
& $6.1$  $^{+0.2}_{-0.1}$ 
& $12.7$ $^{+1.1}_{-0.6}$ 
& $9.5$  $^{+0.7}_{-0.0}$ 
& $-3.4$ $^{-0.8}_{-0.1}$ 
\\[4pt]
\MEPSatNLO,\ \ {\small$\mu=m_{t\bar t}$}
& $4.8$  $^{+0.7}_{-0.3}$ 
& $3.1$  $^{+0.8}_{+0.1}$ 
& $7.9$  $^{+0.5}_{-1.1}$ 
& $5.8$  $^{+0.8}_{-0.4}$ 
& $-7.2$ $^{+0.5}_{-0.4}$ 
\\[3mm]
\MEPS,\ \ {\small$\mu=\mu_\mathrm{QCD}$} 
& $15.0$ $^{+1.9}_{-1.4}$ 
& $11.0$ $^{+1.4}_{-1.1}$ 
& $22.2$ $^{+2.3}_{-2.0}$ 
& $16.6$ $^{+2.2}_{-1.6}$ 
& $-1.1$ $^{+1.7}_{-1.2}$ 
\\[4pt]
\MEPS,\ \ {\small$\mu=m_{t\bar t}$}
& $8.2$  $^{+0.9}_{-0.8}$ 
& $5.9$  $^{+0.6}_{-0.6}$ 
& $12.5$ $^{+1.3}_{-1.2}$ 
& $9.9$  $^{+1.1}_{-1.1}$ 
& $-7.9$ $^{+0.6}_{-0.6}$ 
\\[3mm]
NLO $p\bar{p}\to t\bar{t}$
& $6.0$  
& $4.1$ 
& $9.3$ 
& $7.0$ 
& $-11.1$\\[7pt]\hline\hline
\end{tabular}
\caption{\label{tab:totasym}%
  Top quark forward--backward asymmetry at the parton level.
  We compare experimental data from CDF~\cite{Aaltonen:2012it},
  results from an NLO parton-level $p\bar p\to t\bar t$ calculation
  obtained with MCFM~\cite{Campbell:2010ff,Campbell:2012uf}
  (last row) and predictions in the NLO and LO merging schemes from \Sherpa. 
  The set of uncertainties next to all \Sherpa predictions has been
  determined by varying renormalization and factorization scales in
  the range from $\nicefrac{1}{2}$ (upper) to $2$ (lower).
  We give predictions at the parton level for both of the central
  scale choices discussed in Sec.~\ref{sec:scale_uncertainties}.}
\end{table}

\begin{figure}[t!]
  \centering
  \includegraphics[width=0.47\textwidth]{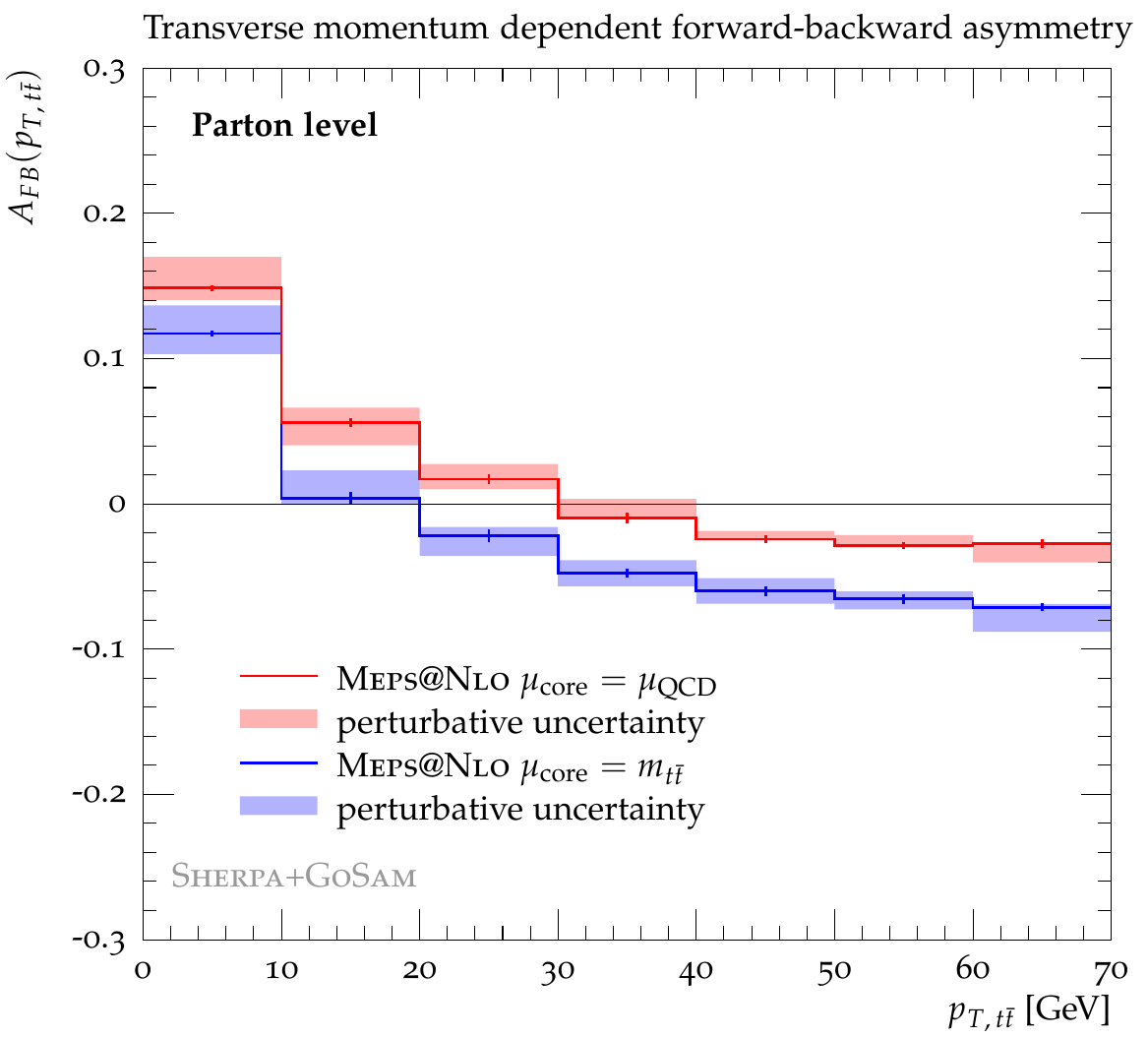}\hfill
  \includegraphics[width=0.47\textwidth]{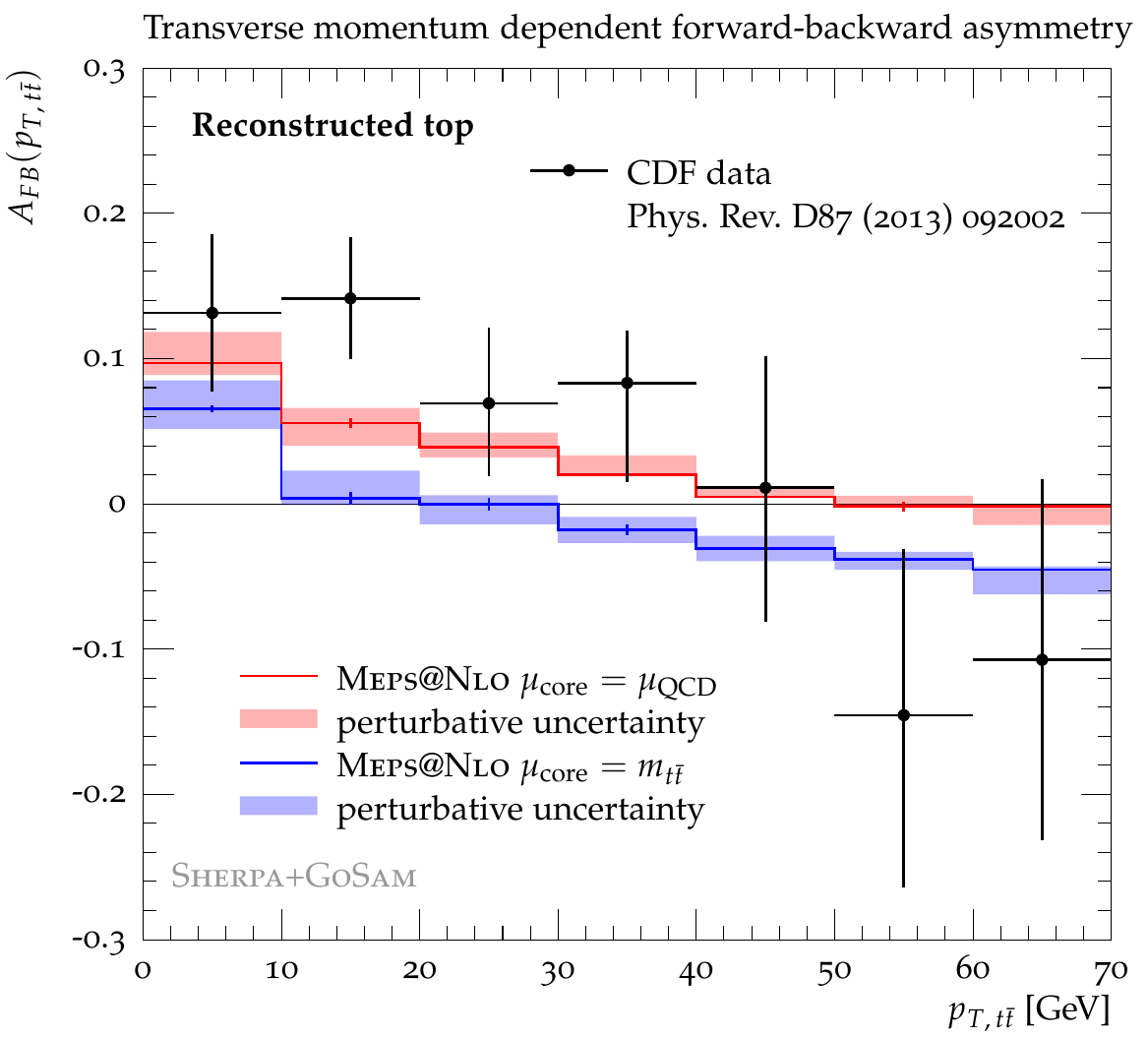}\\
  \vspace*{2mm}
  \includegraphics[width=0.47\textwidth]{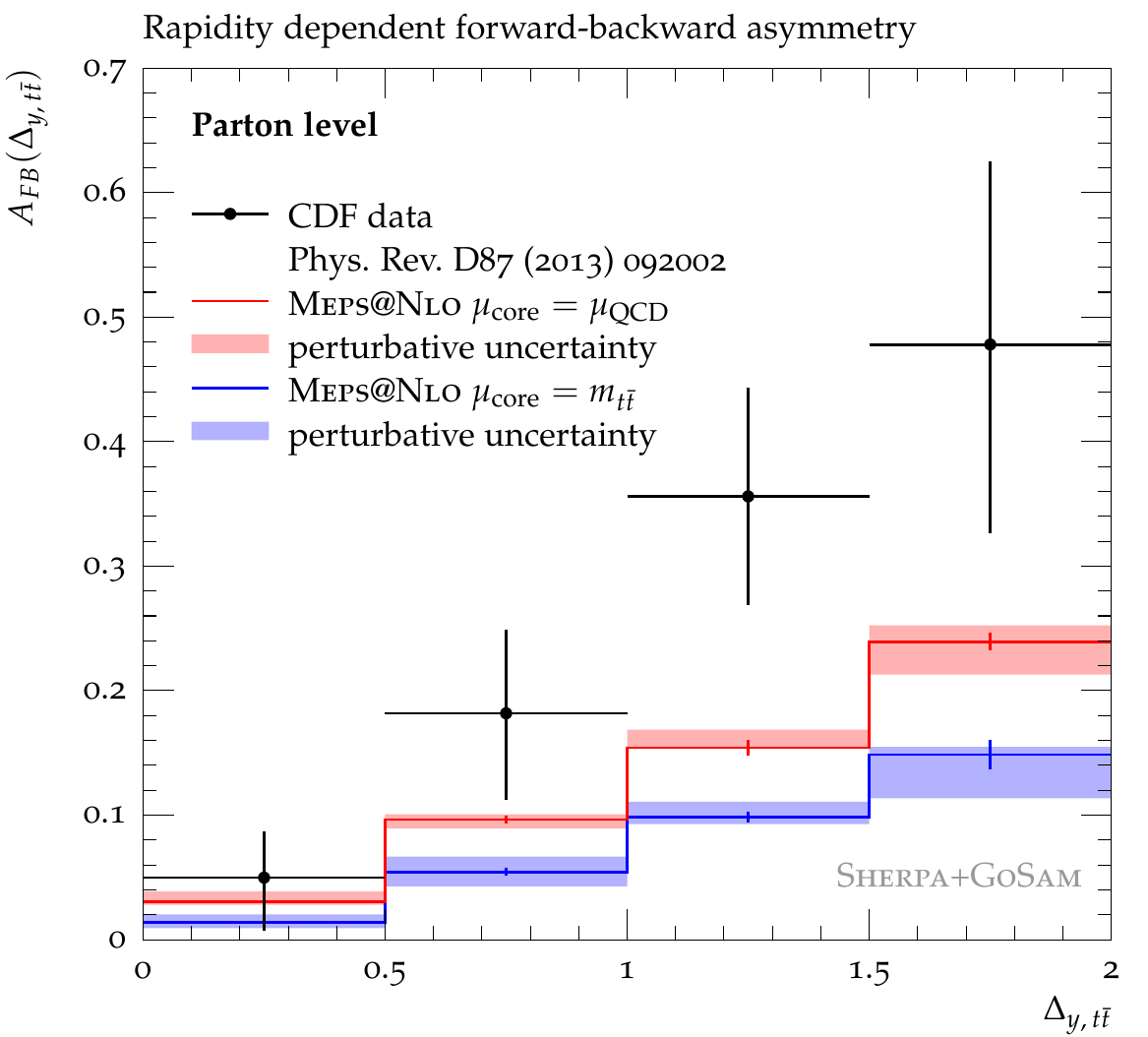}\hfill
  \includegraphics[width=0.47\textwidth]{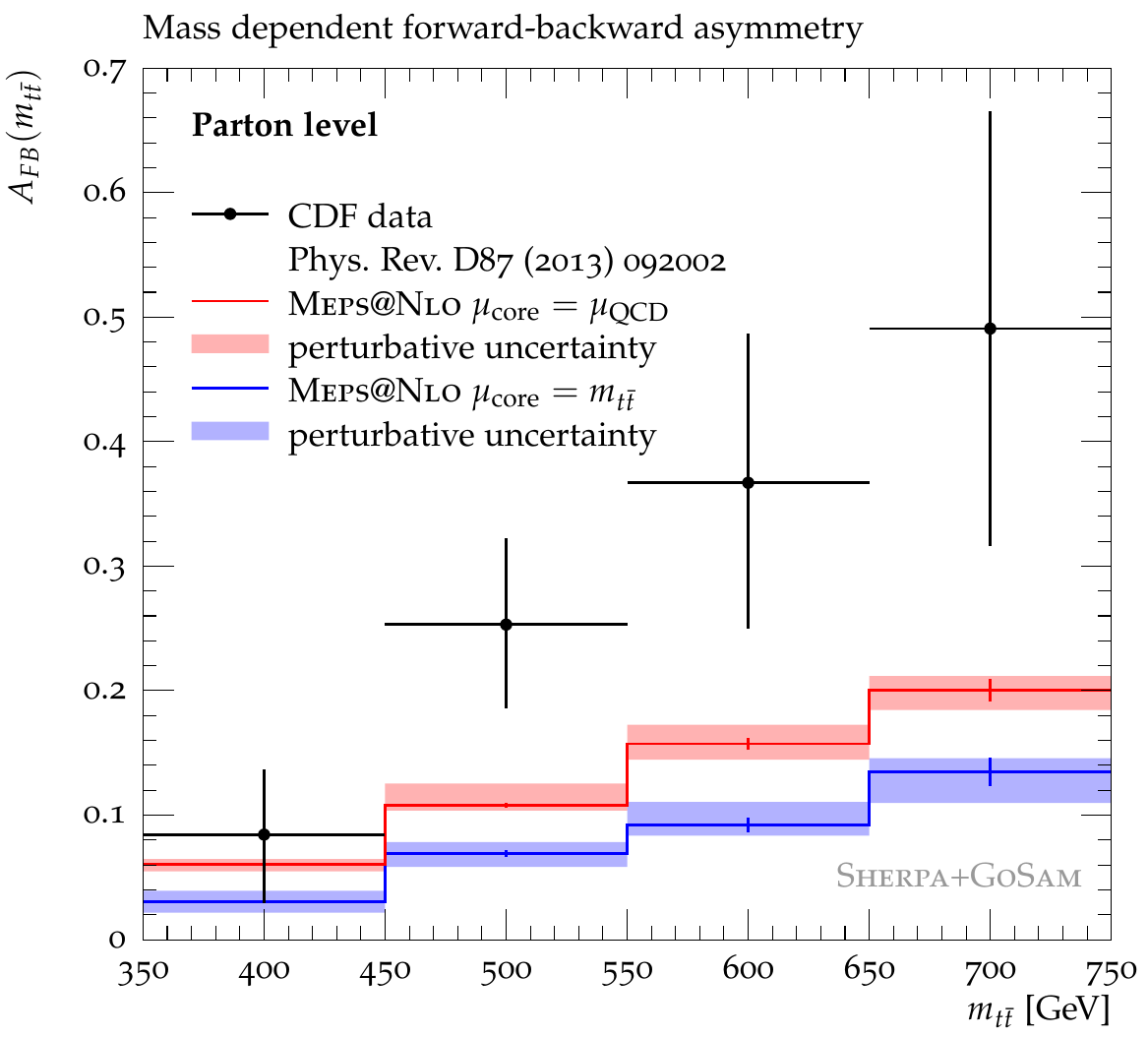}
\caption{\label{fig:bestof}%
  Top quark forward--backward asymmetry in dependence on the transverse 
  momentum (top), the absolute of the rapidity separation,
  $\Delta_{y,t\bar t}\equiv|y_t-y_{\bar t}|$ (bottom left), and the
  invariant mass (bottom right) of the $t\bar t$ system. \MCatNLO zero
  plus one jet merged predictions -- together with their
  uncertainty bands -- are shown for both of the scale choices studied
  in this work, cf.~Sec.~\ref{sec:scale_uncertainties}. The comparison
  is against CDF background subtracted data (top right panel) and against
  parton-level corrected data (bottom panels)~\cite{Aaltonen:2012it}.
  The top left panel shows parton-level results.}
\end{figure}

Table~\ref{tab:totasym} lists the inclusive forward--backward asymmetry,
as well as the asymmetries arising after simple cuts on the invariant mass
of the top-quark pair and its transverse momentum. These cuts separate
the threshold and boosted region in the case of $m_{t\bar t}$,
and the Sudakov and hard-$p_T$ region in the case of $p_{T,t\bar t}$.
We present results for both the \MEPSatNLO and 
the \MEPS methods and compare them to data from the CDF 
collaboration~\cite{Aaltonen:2012it}, and to a fixed-order calculation 
for the asymmetry evaluated at scale $\hat s$ using
\textsc{MCFM}~\cite{Campbell:2010ff,Campbell:2012uf}.

The largest contribution to the overall uncertainty of our predictions
arises from $\mu_\mathrm{R/F}$ variation -- those from other sources are
by and large negligible. We observe a sizable
reduction of scale uncertainties when going from LO to NLO
merging, which was already noted in Sec.~\ref{sec:scale_uncertainties}.
At the same time, however, the central values of $A_{\rm FB}$ decrease
and therefore the discrepancy with the CDF data increases. It should
be stressed that the \MEPS results for $A_{\rm FB}$ have to be
interpreted with caution. The lack of important higher-order corrections
in their calculation, and the correspondingly large scale uncertainties,
point to an agreement with experimental data that is rather accidental. 
Signs of an incomplete, only qualitative description are also given
by the larger spread between the central values associated with the 
different functional forms of the core scale. Moreover, the discrepancy, in
particular for the lowest $p_T$ bin in $A_{\rm FB}(p_{T,t\bar t})$
poses a problem, as can be seen in Fig.~\ref{fig:qcut_scale}.

The disentanglement of the soft and hard regime can be easily achieved
in terms of $p_{T,t\bar t}$. It would therefore be interesting to obtain 
independent measurements for the two different transverse momentum regions,
preferably for an even lower cut. Due to the formal NLO accuracy of
the \MEPSatNLO result for $p_{T,t\bar{t}}>50$\:GeV we expect a better
agreement with data. This is in fact confirmed in Fig.~\ref{fig:bestof}

\subsection{Differential asymmetries}

Figure~\ref{fig:bestof} summarizes our results for the differential
asymmetries. We compare our best predictions, those obtained with
\MEPSatNLO, against the measured distributions for $A_\mathrm{FB}$ in
dependence on the pair transverse momentum, the pair mass and the
absolute rapidity difference between the top quarks. For all
predictions, we show uncertainty bands, which have been obtained
from respectively varying $\mu_\mathrm{R/F}$ and $\mu_Q$ scales by
factors of two and $\sqrt2$, and the merging cut, $Q_\mathrm{cut}$,
from $5$\:GeV to $10$\:GeV where we used $Q_\mathrm{cut}=7$\:GeV for
the central curve. The individual uncertainties are added in
quadrature. They are dominated by the $\mu_\mathrm{R/F}$ variations.
Parton-level to particle-level corrections for the comparison with
the background subtracted data on $A_\mathrm{FB}(p_{T,t\bar t})$
have been computed with \Sherpa in \MCatNLO mode.

We find good agreement with the CDF data for $A_\mathrm{FB}(p_{T,t\bar t})$.
This is an important result, since we obtain this quantitative
agreement in two very different phase-space domains driven by
different physics phenomena: multiple soft and virtual parton emission
in the so-called Sudakov region and hard parton radiation
for larger pair transverse momenta. The prediction based on the QCD
scale choice, which we discussed in Sec.~\ref{sec:scale_uncertainties}, 
gives a slightly better description in the medium pair-$p_T$ range. 
In both cases we observe excellent agreement in the first $p_T$ bin, 
as a result of relying on the subleading-color improved \MCatNLO 
Sudakov exponents.

In the other two observables considered here, the Sudakov region is
spread out over the entire range of the measurement. This leads
to an increase of $A_{\rm FB}$ for larger values of $m_{t\bar t}$ and
$\Delta_{y,t\bar t}$. Both core
scale choices, $\mu_\mathrm{QCD}$ and $m_{t\bar t}$, yield
predictions, which reproduce the linear rise but remain below the
data. Once more, the results obtained using the QCD scale lie closer
to the data, and well in the $2\,\sigma$ range of the given
experimental uncertainty. Note that the asymmetry dependence on these
observables will particularly benefit from the application of
${\cal O}(25\%)$ electroweak corrections, which have not yet been included
in Fig.~\ref{fig:bestof}.

\section{Conclusions}
\label{sec:conclusions}

We have analyzed the top quark forward--backward asymmetry at the Tevatron
collider using a combination of $t\bar{t}$ and $t\bar{t}$+jet calculations
at the next-to-leading order in QCD, merged with a parton shower.

The asymmetry as measured by the CDF and D\O\ collaborations still remains 
a puzzle. While our simulations describe its transverse momentum dependence 
well, the rapidity and mass dependence still show some discrepancies.
More accurate QCD+EW predictions are paramount to clarify whether 
what was measured can be described within the Standard Model, or whether
new physics models are needed. There is also hope that measurements at
the LHC may bring some more insight, although the current situation
still suffers from a lack of analyzed data~\cite{Chatrchyan:2011hk,
  *ATLAS:2012an,*Chatrchyan:2012xv}.

However, a number of interesting points remain: 
firstly, we have achieved a consistent description of both, the Sudakov region 
of the $p_{T,t\bar t}$ spectrum and the high $p_{T,t\bar t}$ domain. 
The transverse momentum dependent forward--backward asymmetry provides 
a first non-trivial test of this method. The fact that it is well described 
by our simulations indicates the potential of the \MEPSatNLO technique. 
Secondly, we demonstrated in a thorough analysis that the application 
of the \MEPSatNLO technique leads to more stable predictions than the \MEPS method. 
It should thus be preferred in experimental analyses.
Thirdly, we showed that including subleading color terms in the first emission 
of the $t\bar{t}$ \MCatNLO has substantial impact on the prediction 
for the asymmetry. This effect cannot simply be subsumed under the standard 
parton-shower uncertainties. The fact that the difference between \MCatNLO 
and \MEPSatNLO predictions is small indicates that the feature remains 
in a parton shower with full color dependence. 
Fourthly, we showed that there is a substantial dependence on the functional
form of the scale, which is used for the $t\bar{t}$ production process.
This dependence is reduced in the \MEPSatNLO method compared to the \MEPS
method, but it indicates that there is still room for explaining the
discrepancy with experiments by QCD+EW corrections at higher orders.

In summary, we have moved one step closer to obtaining an accurate 
inclusive prediction for top-quark pair production at the particle level 
using state-of-the-art Monte Carlo techniques. We used the publicly available 
programs \Sherpa and \GoSam, which makes the results easily reproduceable
and accessible for experiments.

We have performed our calculations at the parton shower level, not including
top-quark decays. Including these decays is important to predict lepton
asymmetries and their correlation with the top quark asymmetry more
reliably, see for example Refs.~\cite{Krohn:2011tw,
  *Falkowski:2011zr,*Berger:2011pu,*Berger:2012nw,*Bernreuther:2012sx,
  *Falkowski:2012cu}.
This analysis is beyond the scope of the present publication, 
and it will be left to future investigation.

\section*{Acknowledgements}

We are indebted to Lance Dixon, Ye Li and Huaxing Zhu for many
fruitful discussions. We thank Francesco Tramontano for help with 
technical aspects of \GoSam. SH's and JH's work was supported by the 
U.S.\ Department of Energy under Contract No.\ DE--AC02--76SF00515.
MS's work was supported by the Research Executive Agency (REA)
of the European Union under the Grant Agreement number
PITN-GA-2010-264564 (LHCPhenoNet).
GL, MS and JW thank the organizers of the Les Houches 2013 workshop 
``Physics at TeV Colliders'' for providing a stimulating atmosphere 
during the final stages of this project.
The work of GL was supported by the Alexander von Humboldt
Foundation, in the framework of the Sofja Kovaleskaja Award Project
``Advanced Mathematical Methods for Particle Physics'', endowed by the
German Federal Ministry of Education and Research.
This research used the CERN computing facilities,
the National Energy Research Scientific Computing Center, which is
supported by the Office of Science of the U.S.\ Department of Energy
under Contract No.\ DE--AC02--05CH11231, and resources provided
by the Open Science Grid, which is supported by the National Science
Foundation and the U.S.\ Department of Energy.

\appendix
\boldmath
\section{IF/FI dipole splitting kinematics and their implications on
  \texorpdfstring{$A_\mathrm{FB}$}{AFB}}\unboldmath
\label{sec:dipole}

In this appendix we will prove that for a parton shower based on
CDST dipole factorization~\cite{Schumann:2007mg} $P_{-+}$ is larger than 
$P_{+-}$, as found by numerical analysis in~\cite{Skands:2012mm}. 

Let us start with the case of initial-state splittings, where the top quark 
plays the role of the spectator parton. Denoting the initial and final state 
momenta before the splitting as $\tilde{p}_{ai}$ and $\tilde{p}_k$, we can
construct the momentum of the top quark after splitting using the variables 
$u_i$ and $x_{ik,a}$ of~\cite{Catani:2002hc} as
\begin{equation}
  p_k\,=\;(1-u_i)\,\tilde{p}_k+u_i\,\left(\frac{1-x_{ik,a}}{x_{ik,a}}
    -\frac{2m_k^2}{Q^2-m_k^2}\right)\,\tilde{p}_{ai}-k_\perp\;,
\end{equation}
where $k_\perp$ is the transverse component, perpendicular to both, $\tilde{p}_k$
and $\tilde{p}_{ai}$. This means, in particular, that $k_\perp$ can be neglected
when analyzing the change of rapidity of the top quark in the splitting process.

We can now easily compute the rapidity difference for the top quark before
and after the splitting:
\begin{equation}
  \Delta y_t\,=\;\frac{1}{2}\ln\left(1
    +\frac{u_i}{1-u_i}\left(\frac{1-x_{ik,a}}{x_{ik,a}}-
      \frac{2m_k^2}{Q^2-m_k^2}\right)
    \frac{\tilde{p}_{ai}^{\,+}}{\tilde{p}_k^{\,+}}\right)
\end{equation}

As $Q^2<0$, the argument of the logarithm is always larger than one, and 
therefore $\Delta y_t$ is positive. This means the top quark is always
pushed in the direction of the momentum of the initial-state quark while 
the anti-top is pushed in the direction of the momentum of the 
initial-state anti-quark. The same is true for the case of final-state
emissions off the top quark with an initial-state spectator, 
as kinematics are defined identically.

At the Tevatron collider, the dominant source of quarks is the proton
beam, while the dominant source of antiquarks is the anti-proton beam.
Therefore, the initial-final and final-initial splittings lead to
$P_{-+}>P_{+-}$.

Final-state splittings with final-state spectator and initial-state
splittings with initial-state spectator do not generate asymmetries,
i.e.\ $P_{-+}=P_{+-}$. Therefore, our argument is complete.

Note that the two different schemes for momentum mapping, which were
compared in~\cite{Carli:2009cg,Hoeche:2009xc}, have similar behavior
with respect to the generation of the asymmetry. This was shown in a
detailed numerical analysis in~\cite{Skands:2012mm}. It can easily 
be explained by the fact that both schemes show a drag of the top
towards larger rapidity due to the color connection with the 
initial-state quark.

It was also shown in~\cite{Skands:2012mm} that a very different momentum
mapping can generate very different asymmetry predictions. This is due to 
the fact that assumptions about the recoil partner being the color-connected
parton in the splitting were relaxed. It is conceivable that, in a
momentum mapping where the recoil is compensated symmetrically by both
initial state quarks, the parton shower does not generate an additional
asymmetry at all. Whether or not this is the more viable physics model
remains to be verified by experiments.

\section{Virtual corrections from \texorpdfstring{\GoSam}{GoSam}}
\label{sec:gosam}

The virtual amplitudes are generated using the
\GoSam~\cite{Cullen:2011ac,*Cullen:2011xs} package, which generates
code for the computation of one-loop integrands. The one-loop
amplitudes are then evaluated at runtime by means of the integrand 
reduction~\cite{Ossola:2006us,*Ellis:2007br} based program 
\Samurai~\cite{Mastrolia:2010nb} and the tensor integral library
\Golem~\cite{Binoth:2008uq,*Cullen:2011kv}. Scalar one-loop integrals
are calculated by \textsc{OneLOop}~\cite{vanHameren:2010cp}.

\begin{table}[t!]
\centering\small
\begin{tabular}{l r r r} \hline \hline \trule
 $d \bar{d} \to t \bar{t} g$  &  \\[2pt]\hline\trule
& \Sherpa+\ \GoSam & Numbers from~\cite{Dittmaier:2008uj} & Universal IR singularity \\[1mm]
$\textrm{Born}$  & $0.5790368001550917\cdot10^{-4}$
                 & $0.5790368001550936\cdot10^{-4}$ \\
$\tilde{c}_{-2}$ & $-5.666666666666674$
                 & $-5.666666666667982$
                 & $-5.666666666666667$ \\
$\tilde{c}_{-1}$ & $-0.7420525970833627$
                 & $-0.7420525970851204$
                 & $-0.7420525970837827$ \\
$\tilde{c}_0$    & $4.912061786537501$
                 & $4.912061774385727$ \\
$c_0$            & $0.2435672441163395$
                 & $0.2435672439083931$ \\[5pt]\hline\trule 
 $g g \to t \bar{t} g$  &  \\[2pt]\hline\trule
& \Sherpa+\ \GoSam & Numbers from~\cite{Dittmaier:2008uj} & Universal IR singularity \\[1mm]
$\textrm{Born}$  & $0.656684336270973\cdot10^{-3}$
                 & $0.656684336270977\cdot10^{-3}$ \\
$\tilde{c}_{-2}$ & $-8.999999999999995$
                 & $-8.999999999455426$
                 & $-9.000000000000002$ \\
$\tilde{c}_{-1}$ & $4.272315663799295$
                 & $4.272315664361962$
                 & $4.272315663817603$ \\
$\tilde{c}_0$    & $16.13909120795238$
                 & $16.13909126125360$ \\
$c_0$            & $0.5295183443224957$
                 & $0.5295183452346090$ \\[5pt]\hline\hline
\end{tabular}
\caption{\label{tab:benchmark_comparison}%
  Numerical results for the benchmark-point comparison
  with~\cite{Dittmaier:2008uj}. The first column contains the numbers
  obtained with the code for the virtual amplitude generated by
  \GoSam. In the second column we report the numbers given
  in~\cite{Dittmaier:2008uj} converted to our normalization. The last
  column contains the coefficients of the poles conform to the
  universal singular behavior derived by
  CDST~\cite{Catani:2002hc}. They are computed using an implementation
  contained in the code for virtual amplitudes.}
\end{table}

We validated the virtual amplitudes from \GoSam with the benchmark points given
in~\cite{Dittmaier:2008uj}, finding full agreement. Since we use a
different normalization compared to~\cite{Dittmaier:2008uj} we define
coefficients $\tilde{c_i}$ ($i=-2,-1,0$) from the coefficients $c_{i}$
given in Eq.~(A.3) of~\cite{Dittmaier:2008uj} as follows:
\begin{equation}
\tilde{c}_{-2}=\frac{c_{-2}}{\frac{\alpha_s}{2\pi}}, \qquad
\tilde{c}_{-1}=\frac{c_{-1}}{\frac{\alpha_s}{2\pi}}, \qquad
\tilde{c}_{0}=\frac{c_{0}}{\frac{\alpha_s}{2\pi}}+\frac{\pi^2}{6}\tilde{c}_{-2}
\qquad \textrm{with} \quad \alpha_s\equiv\alpha_s(m_t)=
0.1075205492734706\,.
\end{equation}
In Tab.~\ref{tab:benchmark_comparison} we then report the details of
the comparison listing all Born numbers and coefficients, which we
found in our calculation.

\section{Color-flow inspired scale choice}
\label{sec:qcd_scale}
In order to quantify the dependence of $A_{FB}$ on renormalization and 
factorization scales, we propose to use two different functional forms
of the scale, where one is insensitive and the other is sensitive to rapidity.
For the former, we select $m_{t\bar{t}}$, while for the latter we select a 
color-flow inspired scale. We call this the ``QCD'' scale for brevity.

The color flow in $q\bar{q}\to t\bar{t}$ / $\bar{q}q\to t\bar{t}$ subprocesses
is unique, hence the QCD scale is identified as
\begin{equation}
  \begin{split}
  \mu_{\rm QCD}^2(q\bar{q}\to t\bar{t})&=2\,p_q p_t=m_t^2-t\;,\\
  \mu_{\rm QCD}^2(\bar{q}q\to t\bar{t})&=2\,p_q p_t=m_t^2-u\;.
  \end{split}
\end{equation}

In the $gg\to t\bar{t}$ subprocess we assign color connections according to 
the method used in~\cite{Marchesini:1987cf}, extended to the case with 
massive final-state quarks. The QCD scale is therefore chosen as
\begin{equation}
  \begin{split}
  \mu_{\rm QCD}^2&(gg\to t\bar{t})=\left\{\begin{array}{ccl}
  m_t^2-t&&w\varpropto\dst\frac{u-m_t^2}{t-m_t^2}+
    \frac{m_t^2}{m_t^2-t}\rbr{\frac{4\,t}{t-m_t^2}+\frac{m_t^2}{s}}\\
  &\text{with weight}&\\
  m_t^2-u&&w\varpropto\dst\frac{t-m_t^2}{u-m_t^2}+
    \frac{m_t^2}{m_t^2-u}\rbr{\frac{4\,u}{u-m_t^2}+\frac{m_t^2}{s}}
  \end{array}\right.\;.
  \end{split}
\end{equation}
Because of the symmetric initial state, $m_t^2-t$ and $m_t^2-u$
are selected with equal probability.

\bibliographystyle{bib/amsunsrt_modp}  
\bibliography{bib/journal}
\end{document}